\documentclass{elsart} 
\usepackage{times,graphicx,amsmath} 

\journal{Nuclear instruments and Methods}

\begin{document}

\renewcommand{\textfraction}{0.1}       
\renewcommand{\floatpagefraction}{0.8}  

\begin{frontmatter}

\vspace*{1cm}

\title{Investigation of vacuum polarization in t-channel radiative Bhabha scattering}

%
\author[label1]{D. Karlen},
\ead{karlen@physics.carleton.ca}
\ead[url]{http://www.physics.carleton.ca/\textasciitilde karlen}
%
\author[label2]{H. Burkhardt},
\ead{Helmut.Burkhardt@cern.ch}
\ead[url]{http://home.cern.ch/hbu/aqed/aqed.html}

\address[label1]{Ottawa Carleton Institute for Physics,Carleton University, Ottawa, Canada K1S 5B6}
\address[label2]{CERN, SL Division, CH-1211 Geneva 23, Switzerland}

\begin{abstract}
We discuss the possibility of a precision measurement of vacuum polarization in
t-channel radiative Bhabha scattering at a high luminosity collider.
For illustration, the achievable precision is estimated for the BaBar experiment at
PEP-II and for the OPAL experiment at LEP.
 
 
\end{abstract}
 

\end{frontmatter}

\section{Introduction}

There is a considerable interest in precision studies of electroweak physics. High energy
data from LEP, SLC and Tevatron probe the electroweak theory on the quantum level.
The radiative corrections are sensitive to the Higgs mass and provide constraints to
theories beyond the standard model \cite{LEPEWWG}.
Very high precision measurements of the anomalous magnetic moment currently
performed at Brookhaven\,\cite{Brown:2001mg} provide another very sensitive
test of the standard model and beyond\,\cite{Czarnecki:2000id,Czarnecki:2001pv}.

A major contribution to the radiative corrections is provided by the running of the
electromagnetic coupling constant $\alpha$ from its value at vanishing
momentum to the effective $Q^2$ of the process studied.
The running of $\alpha$ can be expressed as
\begin{equation}
\alpha(Q^2) = \frac{\alpha(0)}{1-\Delta\alpha_l(Q^2)-\Delta\alpha_{\rm had}(Q^2)} \quad .
\end{equation}
The leptonic contribution $\Delta\alpha_l(s)$ can be calculated with high
accuracy\,\cite{SteinhauserPRLB429}. Theoretical predictions of
the hadronic vacuum polarization have to rely on less well known concepts like
non-perturbative QCD and light quark masses, 
or on experimental uncertainties in the evaluation of the dispersion integral
(the $P$ stands for the principal value)
\begin{equation}
\Delta\alpha_{\rm had}(Q^2) =
- \frac{\alpha\,Q^2}{3\pi}\;\;P\hspace{-1mm}\int_{4m_{\pi}^2}^{\infty}
\frac{R_{\rm had}(s')}{s'(s'-Q^2)} ds' \quad .
\label{eq:ahad}
\end{equation}
$R_{\rm had}$ is the measured QED cross-section of the process e$^+$e$^- \rightarrow $ hadrons,
normalized to the QED cross-section for lepton-pair production.
Eq.\,(\ref{eq:ahad}) applies both to timelike or s-channel
reactions with positive momentum squared ($Q^2 = s > 0$), as well as to spacelike or t-channel
processes with negative momentum squared ($Q^2 = t < 0$).
In the timelike, the integration has a singularity at $s=s'$. This gives
a substantial weight to variations in the cross section close to the actual 
momentum squared of the process. $\Delta\alpha_{\rm had}(s)$ can even become negative for
energies where $R_{\rm had}(s)$ is steeply increasing.
Far from thresholds and resonances, one has
$\Delta\alpha_{\rm had}(Q^2) \approx \Delta\alpha_{\rm had}(-Q^2)$.

\begin{figure}[tp]
  \center\includegraphics[width=12cm]{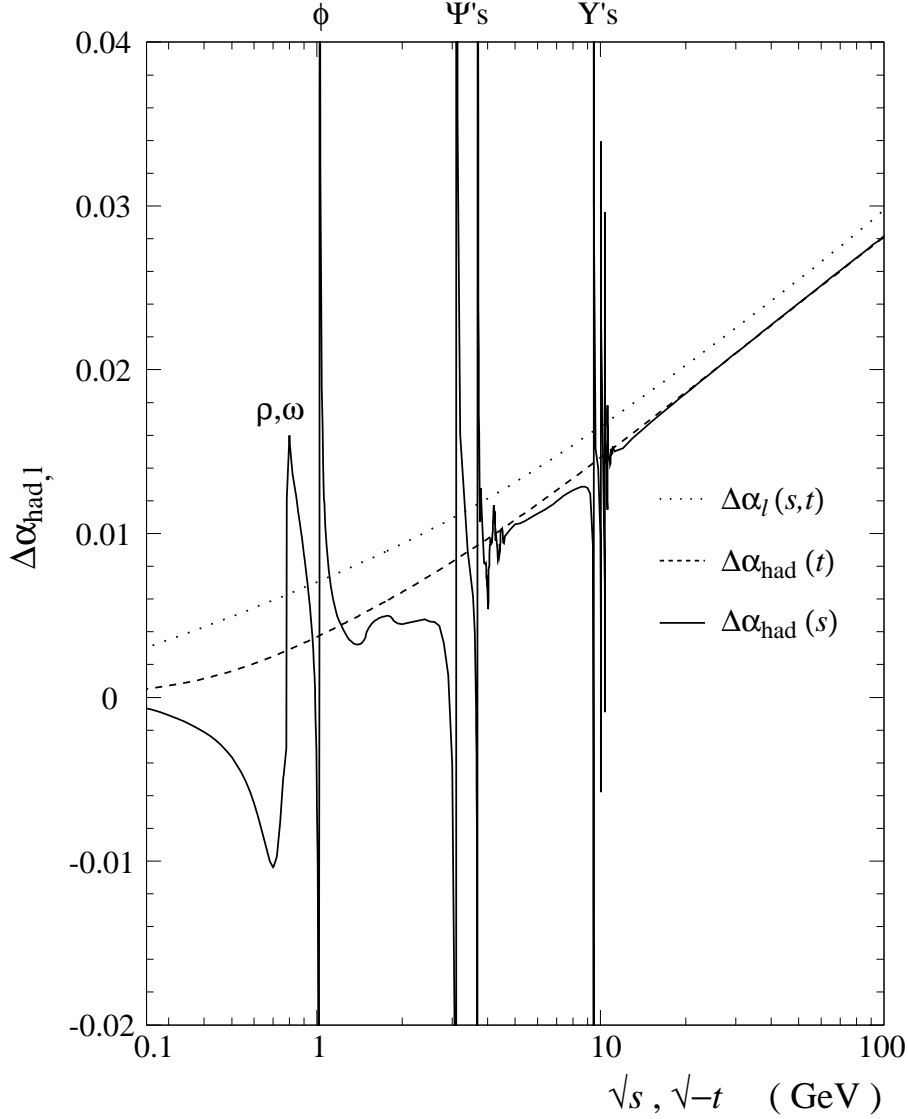}
  \caption{$\protect\Delta\alpha_{l}$  (dotted line) and $\protect\Delta\alpha_{\rm had}$
 at several energies in the $t$ (dashed line) and $s$ channel (solid line).}
  \label{plot:RePiHadLogx}
\end{figure}

For the details of the actual evaluation of the dispersion integral, we refer to
the recently updated work by B.\,Pietrzyk and one of us\,\cite{Burkhardt:2001xp}.
Figure\,\ref{plot:RePiHadLogx} and Table\,\ref{tab:table1}
list numerical values of $\Delta\alpha$
at several values of $\sqrt{s}$ and $\sqrt{-t}$.
The last column gives the leptonic contribution to first order, valid both
for the s and t-channel (higher order corrections are known but actually smaller
than the uncertainty in the hadronic contribution).
Note that the hadronic vacuum polarization is known to about 2\% 
at $t \approx -1\,{\rm GeV}^2$.
This corresponds to relative uncertainties of 0.6\% in $\Delta \alpha$
and $0.9\times10^{-4}$ in $\alpha(1\,{\rm GeV}^2)$.
\begin{table}[tp]\center
\caption{$\Delta\alpha_{\rm had}$ and $\Delta\alpha_{l}$
at several energies in the $t$ and $s$ channel.}
\label{tab:table1}\vskip 1mm
\begin{tabular}{|c|c|c|c|} \hline
$\sqrt{s}$, $\sqrt{-t}$ & $\Delta\alpha_{\rm had}(t)$ & $\Delta\alpha_{\rm had}(s)$ & $\Delta\alpha_{l}(s,t)$ \\
GeV & \% & \% & \% \\ \hline
0.1    & 0.009 $\pm$ 0.001 & -0.009 $\pm$ 0.001 & 0.673 \\
1      & 0.369 $\pm$ 0.009 & -0.638 $\pm$ 0.054 & 1.253 \\
1.5    & 0.524 $\pm$ 0.014 &  0.402 $\pm$ 0.015 & 1.375 \\
10     & 1.463 $\pm$ 0.033 &  0.984 $\pm$ 0.048 & 2.099 \\
91.188 & 2.758 $\pm$ 0.036 & 2.761$\pm$ 0.036   & 3.142 \\
\hline
\end{tabular}
\end{table}

In this paper, we discuss the feasibility of a direct and precise
measurement of the running of $\alpha$ in a single experiment, using radiative Bhabha
scattering.

There are already a number of published papers with rather direct measurements of
the energy dependence in vacuum polarization. Evidence for an observation of 
the hadronic vacuum polarization in the s-channel around the $\phi$ resonance for example
was already reported in 1972\,\cite{Augustin1972}.
More recently, there have been measurements of $\alpha$ at $\sqrt{s} = 58\,{\rm GeV}$ using
muon-pair production\,\cite{Levine:1997zb,Odaka:1998ui}.
Precise measurements of the angular distribution of Bhabha scattering have been interpreted
as further evidence for the running of $\alpha$ in the range of $\sqrt{-t}$ from 
10 to 54~GeV\,\cite{Odaka:1998ui}, and in the ranges $1.5-2.5$~GeV and 
$3.5-58$~GeV\,\cite{Acciarri:2000rx}.

The potential advantage of a measurement based on radiative Bhabha scattering, as proposed
in this paper, is the large cross-section of this process, and the possibility to
cover a large range in $Q^2$ from some GeV down to essentially 0, in a single experiment.

\section{t-channel radiative Bhabha scattering}

\begin{figure}[htbp]
  \center\includegraphics[width=10cm]{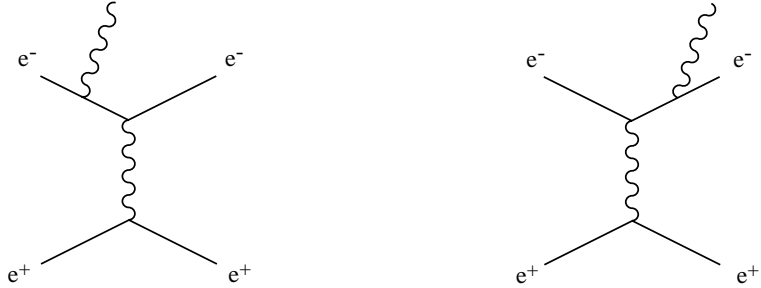}
  \caption{The process of t-channel radiative Bhabha (TRCB) scattering in lowest order.}
  \label{plot:feynbhab.eps}
\end{figure}

The process shown in Fig.\,\ref{plot:feynbhab.eps}, when the exchanged photon is nearly real, can lead to a
distinctive signature in e$^+$e$^-$ collisions, that of a high energy coplanar photon and
electron scattered at wide angles with the other electron scattered at 
a small angle\,\cite{Karlen:1988rd,Karlen:1989wg,Karlen:1987vk}. 
In this note, the process will be referred to as TCRB scattering.
The virtuality, $Q^2$, of the exchanged photon is related to the scattering angle of the electron
on the lower leg. A pole exists in the cross section for nearly zero degree scattering angles.

\begin{figure}[htbp]
  \center\includegraphics[width=\textwidth]{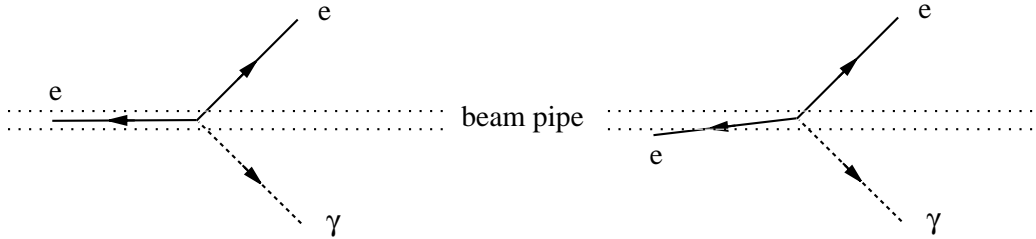}
  \caption{Sketch of the kinematics. On the left, the case of very low $Q^2$ and on the right
  the process with higher $Q^2$ where all particles are detected.}
  \label{plot:SketchKin}
\end{figure}

The event signatures are sketched in Fig.\,\ref{plot:SketchKin} for the two configurations
that are used to measure vacuum polarization.
For events in which the small angle scattered electron goes undetected, the dominant process has
$Q^2$ nearly zero.
When the small angle scattered electron is observed, the $Q^2$ of the exchanged photon depends
on the scattering angle.
The method we are proposing uses the ratio of the number of events in these two
configurations to deduce the running of $\alpha$.

\section{Measurement of TCRB scattering at PEP-II}
As an example, a measurement of TCRB scattering with the BaBar experiment at PEP-II with
500\,fb$^{-1}$ is considered. Collisions take place with a center-of-mass energy of 10.58\,GeV
and the center of mass is boosted with $\beta = 0.49$ in the direction of the electron
(the positive $z$ direction).
Cross sections are calculated for an acceptance that is 
expected to be relatively free of background.
The scattered electron and photon are required to be within the angular acceptance of the
calorimeter $(20^\circ < \theta < 135^\circ)$.
The energies of the photon and wide-angle electron are required
to be at least 2\,GeV and the invariant mass of the pair is to be between 2 and 8\,GeV.
The angles and energies of particles referred to in this section are given in the BaBar laboratory
frame.

The BaBar detector does not include an electron tagger for low backward angles, but it is
envisaged that a detector could be put in place to tag electrons scattered above 300\,mrad
away from the $-z$ axis\,\cite{Babar:1995tdr}.
For the purposes of this study, however, events are classified into
100\,mrad bins of positron scattering angles between 0 and 500\,mrad.
The Belle detector at the KEKB collider has an electron tagger for scattering 
in the angular range of
150 to 300~mrad in the backward direction\,\cite{BelleNimMori}.

The TEEGG event generator\,\cite{Karlen:1987vk} is used to calculate the cross sections of
events in the acceptances define above, and the results are summarized in Table\,\ref{tab:tab1}.
The calculations are done in lowest order, including the $s$ and $t$ channel processes but
without vacuum polarization effects.

\begin{table}[tp]\center\scriptsize
\caption{
Summary of calculations for TCRB scattering for the BaBar acceptances. The first row shows the
lowest order cross sections including the s and t channel contributions, but without vacuum
polarization. The next row shows the number of events that would be collected (in millions)
assuming an integrated luminosity of 500\,fb$^{-1}$. The average $Q^2$ of the events in each bin is
shown to vary from near 0 up to about 1.5\,GeV$^2$. The following row indicates the relative
importance of the t channel contribution, compared to the cross section including all
$s$ and $t$ diagrams. The corrections to $\alpha$ due to leptonic and hadronic loops are calculated
using the repi program \protect\cite{Burkhardt:2001xp}. The average correction values are shown.
The next line indicates the fractional change in cross sections that arise when the effects
of vacuum polarization are included. The last line indicates the relative statistical precision
of the event counts in each bin, which is small compared to the changes in the relative cross
sections in the previous row.
}\label{tab:tab1}\vskip 1mm
\begin{tabular}{|c|c|c|c|c|c|} \hline
& \multicolumn{5}{|c|}{Range of e$^+$ scattering angles (rad)} \\ \cline{2-6}
& $\theta_{e^+} < .1$ & $.1 < \theta_{e^+} < .2$ & $.2 < \theta_{e^+} < .3$ & $.3 < \theta_{e^+} < .4$ & $.4 < \theta_{e^+} < .5$ \\ \hline
$\sigma(\alpha(0))$ (pb)                                           & 222     & 22    & 13.2  & 10.0  & 8.3   \\
$N$, (in $10^6)$                                                   & 111     & 11    & 6.6   & 5.0   & 4.1   \\
$<Q^2>$ (GeV)$^2$                                                  & 0.005   & 0.15  & 0.44  & 0.89  & 1.51  \\
$1-\dfrac{\sigma(t)}{(\sigma(s+t)}$, (in $10^{-4}$)                & 0.6     & 52    & 130   & 255   & 415   \\
$<\Delta\alpha_l>$, (in $10^{-4}$)                                 & 34      & 99    & 115   & 125   & 133   \\
$<\Delta\alpha_{\rm had}>$, (in $10^{-4}$)                         & 0.4     & 11    & 23    &  35   &  44   \\
$\dfrac{\sigma(\alpha(Q^2))}{\sigma(\alpha(0))}-1$, (in $10^{-4}$) & 69      & 223   & 281   & 327   & 365   \\
$1/\sqrt{N}$, (in $10^{-4}$)                                       &  1.0    & 3.0   & 3.9   & 4.5   & 4.9   \\
\hline
\end{tabular}
\end{table}

\begin{figure}[tp]
  \center\includegraphics[width=\textwidth]{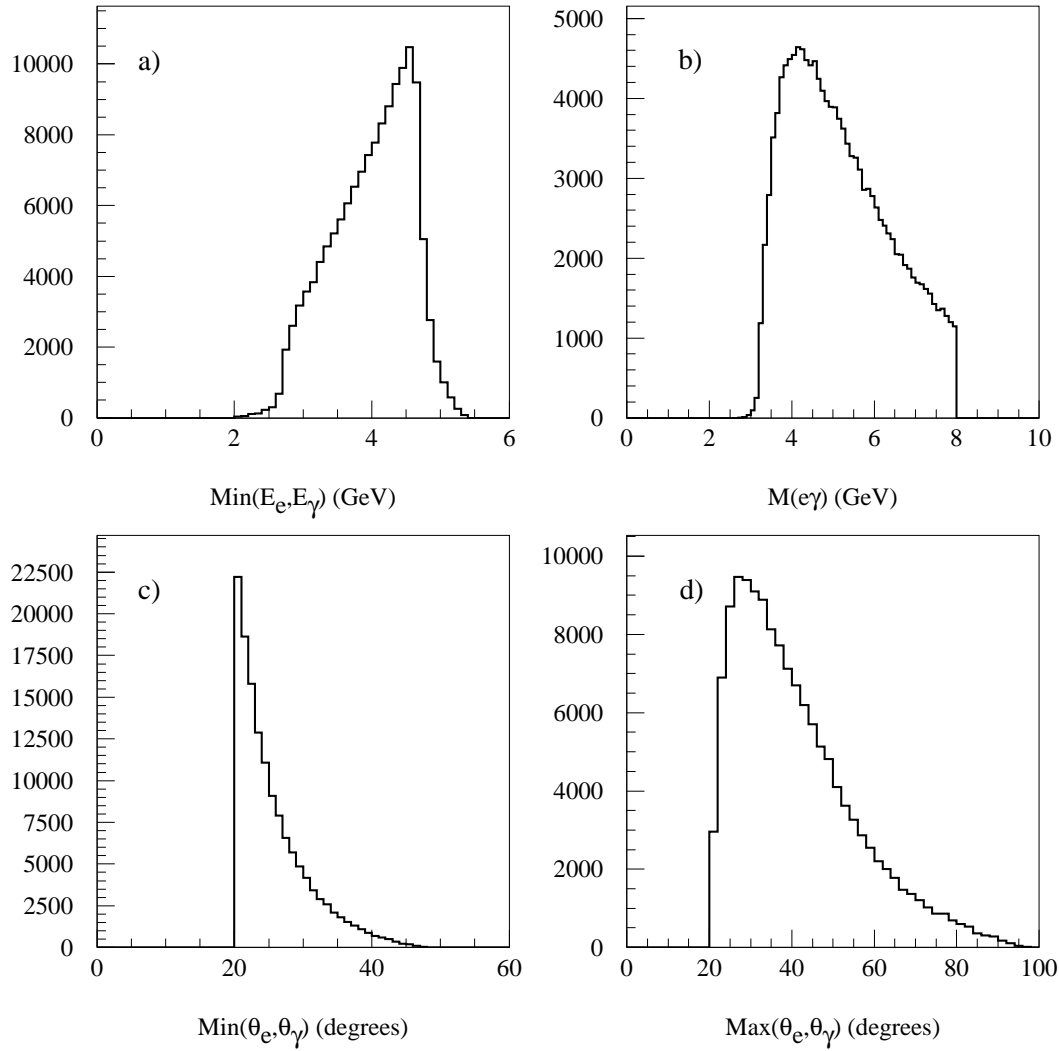}
  \caption{Some distributions of TCRB scattering events in the BaBar acceptance
  for  $\theta_{e^+} < 0.5$\,rad.
  a) The minimum energy of the photon and wide angle electron (required to be
   above 2\,GeV). The natural cutoff above about 2.5\,GeV is due to momentum
    conservation in the 3 body final state.
  b) The e$\gamma$ invariant mass. This is required to be above 2\,GeV, in order to
  reduce background from various processes, and below 8\,GeV to
  reduce contamination from 2$\rightarrow$2 processes, which would appear at the
  center-of-mass energy.
  c) The minimum scattering angle of the photon and wide angle electron
   (required to be above 20$^\circ$).
  d) The maximum scattering angle of the photon and wide angle electron
    (required to be below 135$^\circ$).}
  \label{plot:DisBabar}
\end{figure}

\begin{figure}[tp]
  \center\includegraphics[width=12cm]{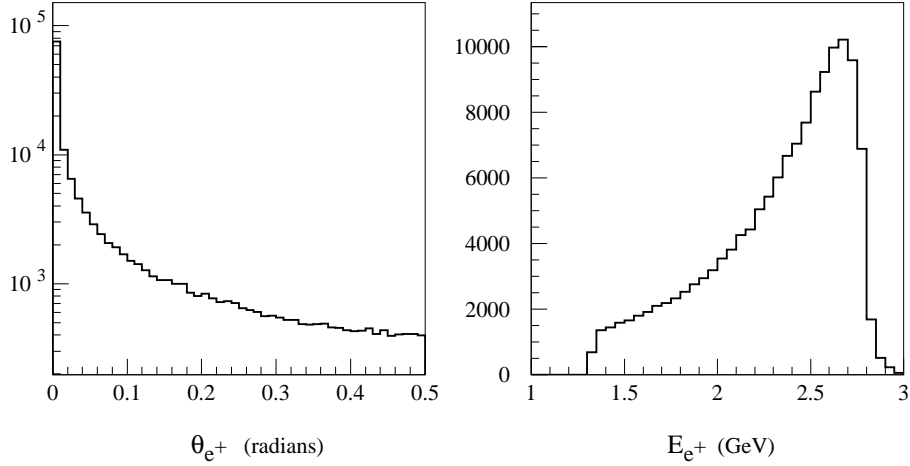}
  \caption{Distributions of the low angle scattered positron.
  a) The scattering angle with respect to the -z axis.
  b) The energy of the positron.}
  \label{plot:AngDis}
\end{figure}

\begin{figure}[tp]
  \center\includegraphics[width=\textwidth]{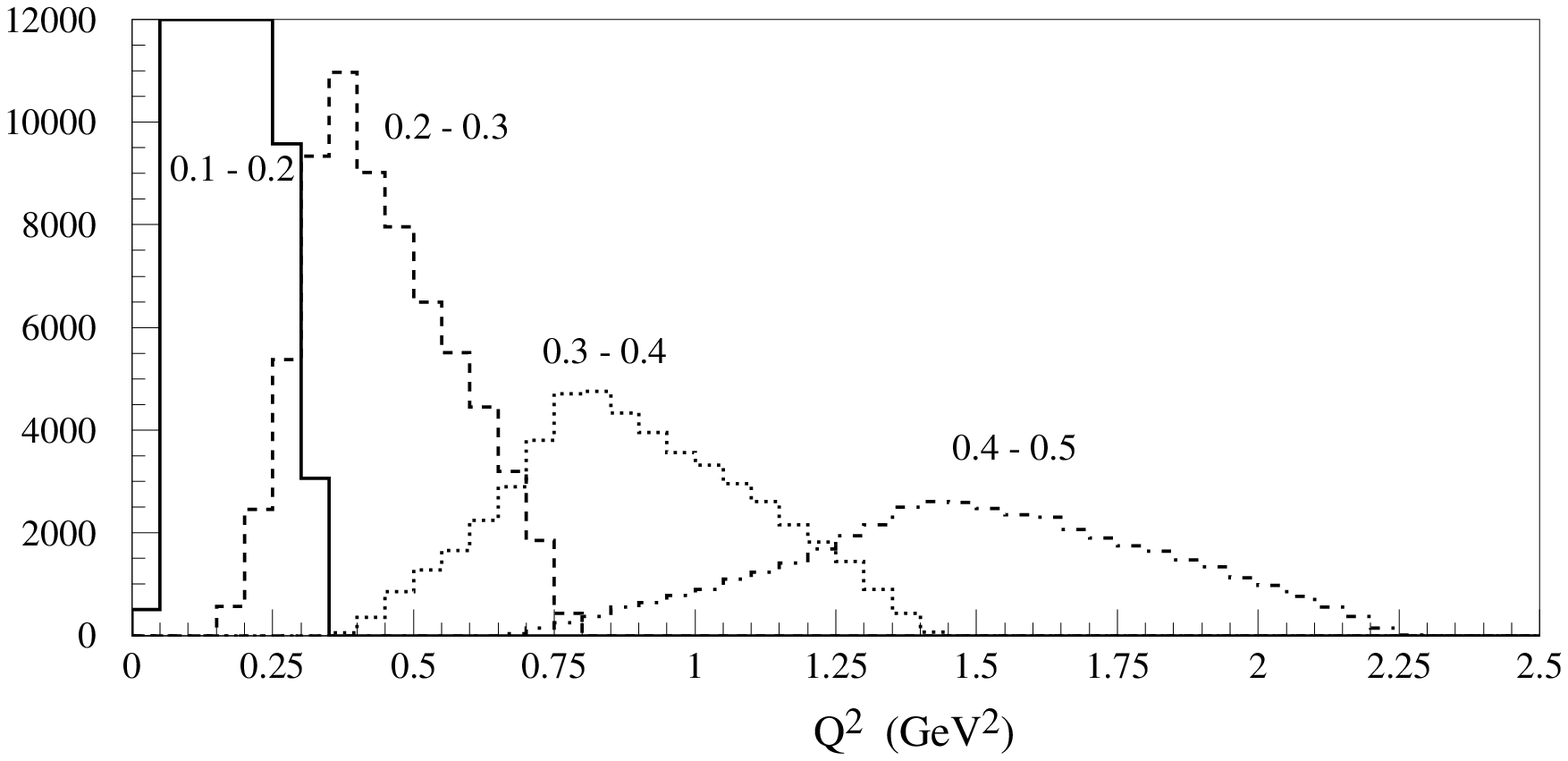}
  \caption{The distributions of $Q^2$ are shown for the different bins in scattering
   angle of the low angle positron. The solid histogram is $0.1 < \theta_{e^+} < 0.2$;
   the dashed is $0.2 < \theta_{e^+} < 0.3$, the dotted is $0.3 < \theta_{e^+} < 0.4$,
   and the dot-dashed is $0.4 < \theta_{e^+} < 0.5$ (all in radians).}
  \label{BabarQ2.eps}
\end{figure}

Some characteristics of the events inside the full acceptance, $\theta_{e^+} < 0.5$\,rad,
are shown in Fig.\,\ref{plot:DisBabar}. The bulk of events are well away from the edges of
most of the acceptance criteria, apart from the requirement on the minimum
scattering angle of the electron and photon. Distributions of the low angle
positron are shown in Fig.\,\ref{plot:AngDis}.

Vacuum polarization has different strength in the different bins of positron
scattering angle due to the fact that the bins involve different values of $Q^2$.
Since the TCRB scattering process is dominated by the $t$ channel contribution,
the modification to the cross section due to vacuum polarization can be approximated by,

\begin{equation}
\sigma\left(\alpha(Q^2)\right) = \frac{\alpha^2(Q^2)}{\alpha^2(0)} \ \ \sigma\left(\alpha(0)\right).
\end{equation}

In order to measure the variation of $\alpha(Q^2)$ in a simple fashion, the ratio of event
counts in each bin to that in the lowest angle bin can be used. In this way
systematic uncertainties arising from detector acceptance
can be reduced. Figure\,\ref{BabarQ2.eps} shows the $Q^2$ distributions for the four samples above
100\,mrad.

The event ratios,
$R_i = \sigma(\theta_{{\rm low},i} < \theta_{e^+} < \theta_{{\rm high},i}) \,/ \sigma(\theta_{e^+} < 0.1)$,
can be measured and compared to the expected ratios in absence of vacuum polarization effects,
$R_i(\alpha(0))$. 
Deviations in the double ratios, $\Delta D_i = R_i / R_i(\alpha(0)) -1$,
measure the strength of vacuum polarization. 
From the last row in Table\,\ref{tab:tab1} one can see that the
the relative statistical precision of the event rates is $3-5$ $\times 10^{-4}$,
which corresponds to relative statistical
uncertainties in $\alpha(Q^2)$ of about $2\times 10^{-4}$.
This is still about twice as large as the present relative 
uncertainty in the prediction of $\alpha(Q^2)$, for $Q^2$ near 1~GeV$^2$.

\section{Systematic uncertainties}
The statistical precisions of the event ratios are at the level of a few parts
in 10$^4$. It is a significant challenge to limit systematic effects in the
measurement of event ratios to the same level. In this section, some of
these effects are considered and are summarized in Table\,\ref{tab:tab2}.

\begin{table}[tp]\center\scriptsize
\caption{Summary of systematic uncertainties in the measurement of event ratios.
The first row indicates the deviation in the double ratio from 1, due to
vacuum polarization. The second row indicates the statistical uncertainty
in the double ratios for each bin. The next rows show the deviations in the
double ratio $D_i$ for each bin for a number of situations that can result in
systematic uncertainties. Note that the systematic uncertainties are correlated.
}\label{tab:tab2}\vskip 1mm
\begin{tabular}{|c|c|c|c|c|} \hline
& \multicolumn{4}{|c|}{Range of e$^+$ scattering angles (rad)} \\ \cline{2-5}
& $.1 < \theta_{e^+} < .2$ & $.2 < \theta_{e^+} < .3$ & $.3 < \theta_{e^+} < .4$ & $.4 < \theta_{e^+} < .5$ \\ \hline
$\Delta D_i$ [vac. pol.] (in $10^{-4}$)                                    & 154 & 211 & 257 & 294 \\
Statistical uncertainty  (in $10^{-4}$)                                    &   3 &   4 &   5 &   5 \\
$\Delta D_i$ [min$(\theta_e$,$\theta_\gamma) > 20.2^\circ$] (in $10^{-4}$) &  10 &   7 &   7 &   9 \\
$\Delta D_i$ [min$($E$_e,$E$_\gamma) > 2.04$ GeV] (in $10^{-4}$)           &   0 &   0 &   1 &  27 \\
$\Delta D_i$ [$M_{e\gamma}^2 < 62$ GeV$^2$] (in $10^{-4}$)                 &  12 &  15 &   3 &   1 \\
$\Delta D_i$ [$\Delta\theta_{e^+}$ = 1 mrad] (in $10^{-4}$)                &  77 &  57 &  26 &  30 \\
$\Delta D_i$ [radiative corrections] (in $10^{-4}$)                        &  65 &  23 & 100 & 142 \\
\hline
\end{tabular}
\vspace{5mm}
\end{table}

\subsection*{Calorimeter angular acceptance}
Figure\,\ref{plot:DisBabar}c shows that the events are very forward peaked. The number of accepted
events is therefore very sensitive to the location of the cut on the minimum\
scattering angle. The shape of this distribution depends somewhat on $Q^2$, so that
the double ratios have a residual dependence on the precise location of the cut.
For example, if the cut is moved to 20.2$^\circ$, the double ratio changes by a
factor of $(10\times10^{-4})$. The values are shown for each bin in Table\,\ref{tab:tab2}.

\subsection*{Energy scale of the calorimeter}

Modifying the minimum energy cut from 2\,GeV to 2.04\,GeV, results in a change in
the double ratio of events for the larger $Q^2$ bins. Changing the maximum invariant
mass from 8 to 7.9\,GeV affects the double ratio most significantly for lower $Q^2$
bins.

\subsection*{Low angle tagger acceptance}
As expected, the measurement requires a precise knowledge of the angular
acceptance of the small angle scattered positrons. Table\,\ref{tab:tab2}
shows the change in the
double ratio that results in a change in all angular bins by 1\,mrad. If the edges
of the angular acceptance can be known to a precision of 0.1\,mrad, the systematic
error will be about the same magnitude as the statistical error.
Additional corrections due the beam spot size and position in the
interaction region and due to the beam divergence will have to be taken into account.
They are generally well known on average, from the knowledge of the machine optics and
monitoring of beam sizes and positions, such that the contribution
to the systematic uncertainty will be small.

\subsection*{Radiative corrections}
The TEEGG program implements photonic radiative corrections to the process of
TCRB scattering using the equivalent photon approximation. The effect of these
radiative corrections modifies the double ratio by significant amounts.
The magnitude of the radiative correction needs to be known at the level of 10\%
or better, for this not to lead to a dominant systematic uncertainty. A better
understanding of the sensitivity of radiative corrections requires a simulation
of the experiment and perhaps a more exact calculation.

\subsection*{Background processes}
Background events are expected to come from Bhabha scattering (when one electron
undergoes bremsstrahlung in the detector material) and two photon final states
(when one photon converts). These backgrounds are reduced by the cut on the\
invariant mass. When these events are accompanied by initial state radiation, the
invariant mass cut is no longer effective.
Such events however do not exhibit the strong charge asymmetry of the wide-angle
scattered electron, and thus the size of the remaining background can be measured
directly in the data sample.

Another background source will be due to off-momentum particles lost from
the beam. As off-momentum particles originate in beam-gas scattering, 
good vacuum conditions
in the straight sections around the experiment will be important.
The background rates due to off-momentum background particles reaching the detectors
can be monitored by analyzing random beam crossing samples.
This source of background can be distinguished from the signal through the use of
kinematic constraints imposed by four-momentum conservation.

\section{Measurement of TCRB scattering at LEP}

The integrated luminosity recorded at LEP is three orders of magnitude less than that
expected for PEP II. In addition, the cross section for TCRB scattering is lower
for a fixed acceptance at higher energies. Nevertheless, it is possible to make
an interesting measurement using the LEP data set.

The cross section for TCRB scattering at LEP energies for configurations where
both the electron and photon are scattered into the central detector is small,
typically a few pb. By requiring instead that the photon scatters into the forward
detector (used for precision luminosity measurements of Bhabha scattering),
the cross section increases by more than a factor of 30. 

The set of requirements considered for LEP1 and LEP2 are summarized below.
The electron is required to be scattered in the forward hemisphere ($0 < \cos\theta < 0.9$) and a forward
detector tag (due to the photon with $25 < \theta_\gamma < 58$~mrad ) is required in the same hemisphere.
The minimum transverse momentum of the wide angle scattered electron is set to 1.1~GeV for LEP1 and
2.4~GeV at LEP2.
The minimum energy of the tagging photon is 30 (70)~GeV for LEP1 (LEP2). 

The TEEGG generator is used to evaluate the cross sections in these acceptances.
Tables\,\ref{tab:tab3} and \ref{tab:tab4} summarize the measurements at LEP1 and LEP2, assuming integrated
luminosities of 100 and 400 pb$^{-1}$, respectively. In the lower two rows
of the last column of these tables, it is seen that the running of
the fine structure constant modifies the cross section by about twice the
statistical uncertainty for both cases.

\begin{table}[tp]\center\scriptsize
\caption{Summary of calculations for TCRB scattering for the LEP acceptances.
See the caption for Table\,\ref{tab:tab1} for an explanation of the entries.}\label{tab:tab3}\vskip 1mm
\begin{tabular}{|c|c|c|c|c|c|} \hline
& \multicolumn{5}{|c|}{Range of e$^+$ scattering angles (mrad)} \\ \cline{2-6}
& $\theta_{e^+} < 25$ & $25 < \theta_{e^+} < 36$ & $36 < \theta_{e^+} < 47$ & $47 < \theta_{e^+} < 58$ & $25 < \theta_{e^+} < 58$ \\ \hline
$\sigma(\alpha(0))$ (pb)                              & 486          & 96    & 66    & 55    & 217   \\
$N$, (in $10^3)$                                      & 49           & 9.6   & 6.6   & 5.5   & 22    \\
$<Q^2>$ (GeV)$^2$                                     & 0.042        & 1.91  & 3.59  & 5.77  & 3.40  \\
$1-\dfrac{\sigma(t)}{(\sigma(s+t)}$, (in $10^{-2}$)   & 0.004        & 0.36  & 0.63  & 0.89  & 0.58  \\
$<\Delta\alpha_l>$, (in $10^{-2}$)                    & 0.30         & 1.37  & 1.48  & 1.56  & 1.45  \\
$<\Delta\alpha_{\rm had}>$, (in $10^{-2}$)            & 0.02         & 0.49  & 0.62  & 0.73  & 0.60  \\
$\dfrac{\sigma(\alpha(Q^2))}{\sigma(\alpha(0))}-1$, 
 (in $10^{-2}$)                                       & 0.7          & 3.8   & 4.3   & 4.8   & 4.2   \\
$1/\sqrt{N}$, (in $10^{-2}$)                          & 0.5          & 3.2   & 3.9   & 4.3   & 2.1   \\
\hline
\end{tabular}
\vspace{5mm}
\end{table}

\begin{table}[tp]\center\scriptsize
\caption{Summary of calculations for TCRB scattering for the LEP2 acceptances.
See the caption for Table\,\ref{tab:tab1} for an explanation of the entries.}\label{tab:tab4}\vskip 1mm
\begin{tabular}{|c|c|c|c|c|c|} \hline
& \multicolumn{5}{|c|}{Range of e$^+$ scattering angles (mrad)} \\ \cline{2-6}
& $\theta_{e^+} < 25$ & $25 < \theta_{e^+} < 36$ & $36 < \theta_{e^+} < 47$ & $47 < \theta_{e^+} < 58$ & $25 < \theta_{e^+} < 58$ \\ \hline
$\sigma(\alpha(0))$ (pb)                              & 110          & 20.1  & 14.0  & 11.6  & 46    \\
$N$, (in $10^3)$                                      & 44           & 8.0   & 5.6   & 4.6   & 18    \\
$<Q^2>$ (GeV)$^2$                                     & 0.19         & 9.0   & 16.9  & 27.2  & 16.1  \\
$1-\dfrac{\sigma(t)}{(\sigma(s+t)}$, (in $10^{-2}$)   & 0.004        & 0.36  & 0.63  & 0.89  & 0.58  \\
$<\Delta\alpha_l>$, (in $10^{-2}$)                    & 0.37         & 1.64  & 1.76  & 1.85  & 1.73  \\
$<\Delta\alpha_{\rm had}>$, (in $10^{-2}$)            & 0.05         & 0.84  & 1.00  & 1.12  & 0.96  \\
$\dfrac{\sigma(\alpha(Q^2))}{\sigma(\alpha(0))}-1$, 
 (in $10^{-2}$)                                       & 0.8          & 5.1   & 5.7   & 6.2   & 5.6   \\
$1/\sqrt{N}$, (in $10^{-2}$)                          & 0.5          & 3.5   & 4.2   & 4.7   & 2.3   \\
\hline
\end{tabular}
\end{table}

\begin{figure}[tp]
  \center\includegraphics[width=12cm]{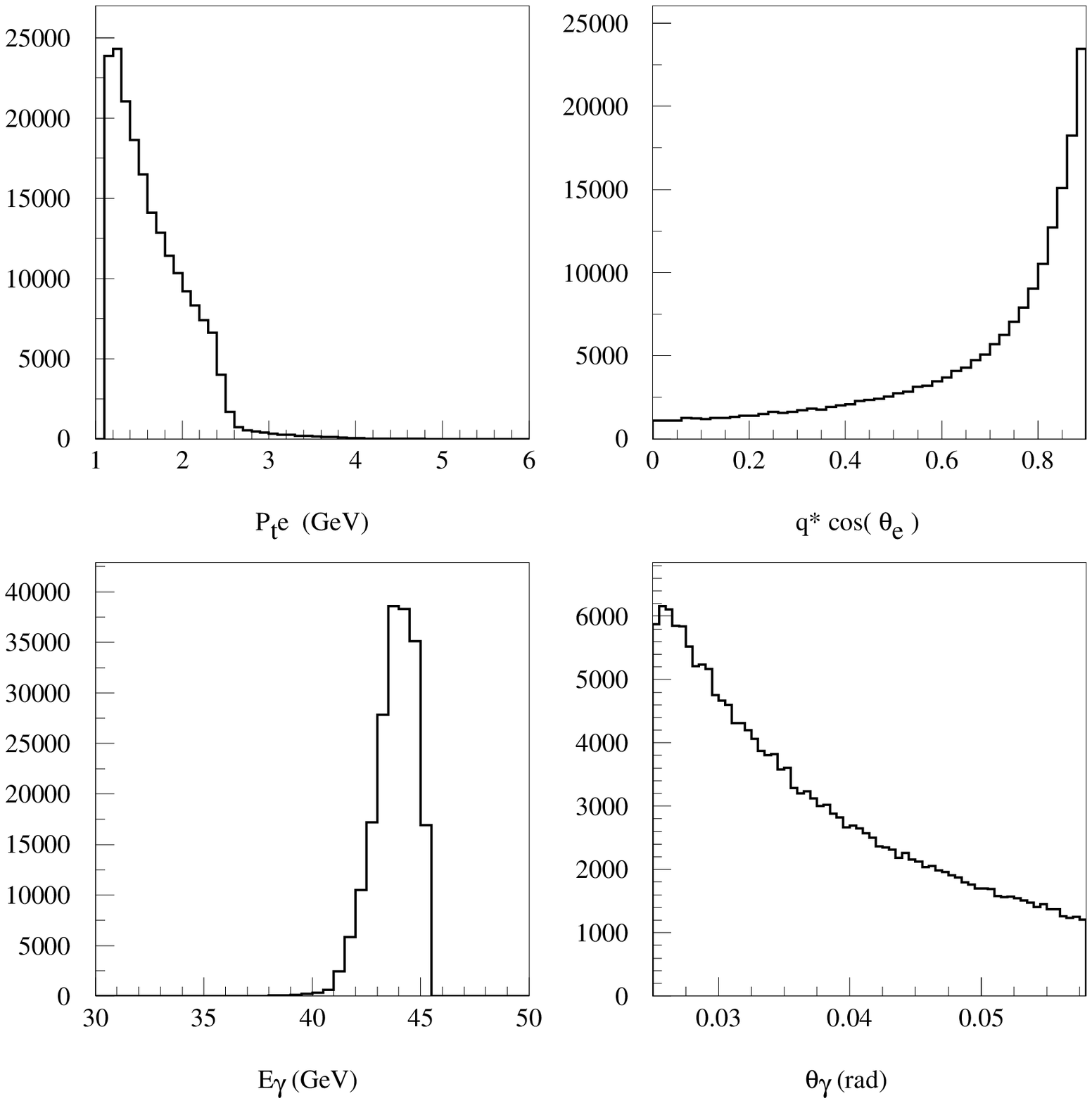}
  \caption{Distribution for the LEP1 acceptance.}
  \label{plot:DisLep1}
\end{figure}

\begin{figure}[tp]
  \center\includegraphics[width=12cm]{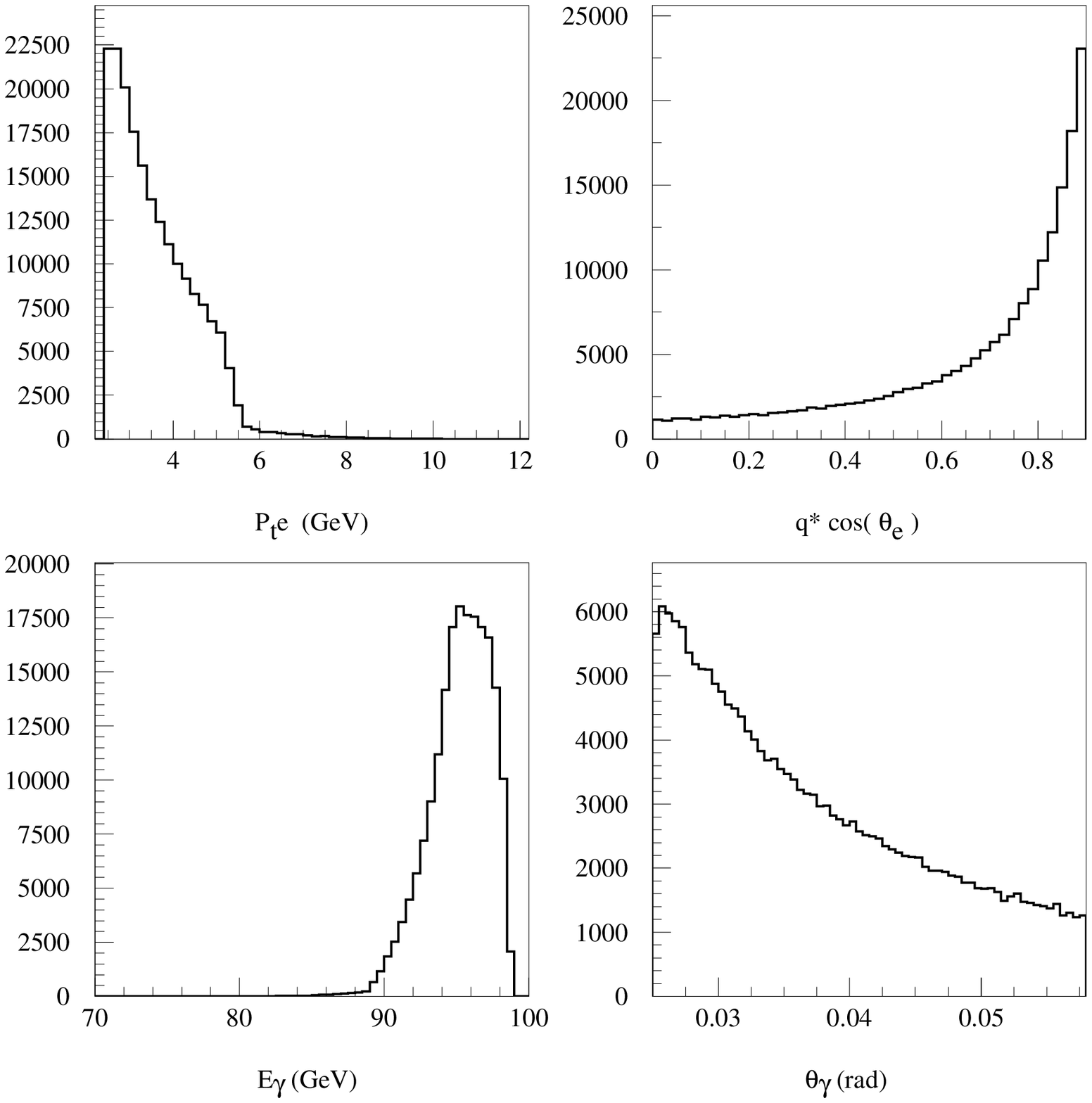}
  \caption{Distribution for the LEP2 acceptance.}
  \label{plot:DisLep2}
\end{figure}

Some characteristics of the events are shown in Figs.\,\ref{plot:DisLep1} and \ref{plot:DisLep2}.
Systematic uncertainties will be smaller than the statistical uncertainties provided
the photon angular acceptance is known with 0.1\,mrad, and the scale of the
transverse momentum of the wide scattered electron is known within 50 and 100 MeV
for LEP1 and LEP2 respectively.

\section{Summary}

\begin{figure}[tp]
  \center\includegraphics[width=12cm]{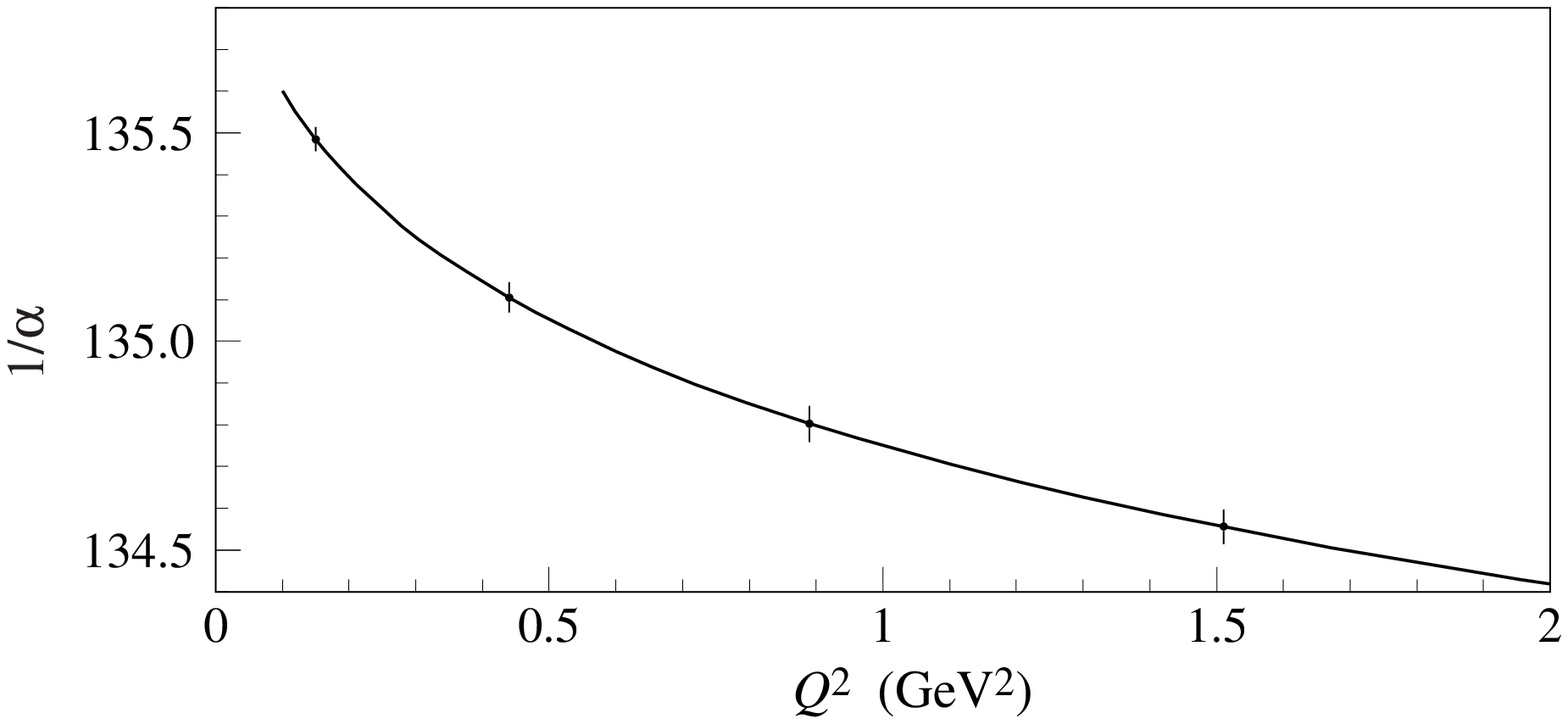}
  \caption{The potential measurement of the running of $\alpha$ at PEP-II.
   The error bars are statistical only.}
  \label{plot:PotentialRunningMeas}
\end{figure}

A precise measurement of vacuum polarization for $Q^2 < 1.5\,{\rm GeV}^2$ could be
possible at the asymmetric B factory, PEP-II (with an improvement in precision by an
order of magnitude compared to previous measurements).
A plot of the running of the fine structure constant that could result from such
a measurement is shown in Fig.\,\ref{plot:PotentialRunningMeas}. 
While the relative statistical precision of approximately $2\times 10^{-4}$ is
somewhat larger than the present uncertainties in the calculations of
$\alpha(Q^2)$, such a measurement could still give an impressive illustration of the running
of one of the most fundamental  constants in nature.

With the data sample recorded at LEP it is possible to use this approach
to confirm, at the few standard
deviations level, that the fine structure constant runs in the
range $\sqrt{-t}$ from near 0 to 5~GeV.

It may be possible to apply the technique proposed here
at the KEKB collider or at other high luminosity electron positron colliders. 
In addition, approaches that have already been applied at Tristan and LEP,
using Bhabha scattering and muon pair production, can be applied at the
high luminosity colliders to further improve the precision on the determination of vacuum polarization,
and to extend the range in $Q^2$.

The authours would like to thank W.\,Kozanecki for useful discussions.


\clearpage 

\bibliographystyle{utcaps}
\bibliography{BhaVac}

\end{document}